\documentclass[useAMS,usenatbib,usedcolumn]{mn2e}

\input ifpdf.sty

\ifpdf
 \usepackage[pdftex,colorlinks,urlcolor=blue,draft]{hyperref}
\else
 \usepackage{hyperref}
\fi

\usepackage{graphicx}
\usepackage{relsize}
\usepackage{amsbsy}
\usepackage{amssymb}
\usepackage{amsmath}

\usepackage{colortbl}

\newcommand{\unit}[1]{\,\mathrm{#1}}

\newcommand{\oAnd}{{\scriptscriptstyle \rm M31}}
\newcommand{\oMW}{{\scriptscriptstyle \rm MW}}

\newcolumntype{d}[0]{D{.}{.}{1}}

\newlength{\wminus}
\begin{document}
\title[Discs of Satellites: the new dwarf spheroidals]{Discs of Satellites: the new dwarf spheroidals}

\author[M. Metz, P. Kroupa and H. Jerjen]
{Manuel Metz$^{1}$\thanks{E-mail: \url{mmetz@astro.uni-bonn.de}}, Pavel Kroupa$^{1}$ and Helmut Jerjen$^{2}$\\
$^1$Argelander-Institut f\"ur Astronomie, Universit\"at
Bonn, Auf dem H\"ugel 71, D--53121 Bonn, Germany\\
$^2$Research School of Astronomy and Astrophysics, ANU, Mt. Stromlo Observatory, Weston ACT 2611, Australia
}

\date{Received 2008 November 6 / Accepted 2009 January 12}
\pagerange{\pageref{firstpage}--\pageref{lastpage}} \pubyear{2009}

\maketitle
\label{firstpage}

\begin{abstract}
The spatial distributions of the most recently discovered ultra faint dwarf satellites around the Milky Way and the Andromeda galaxy are compared to the previously reported discs-of-satellites (DoS) of their host galaxies. In our investigation we pay special attention to the selection bias introduced due to the limited sky coverage of SDSS. We find that the new Milky Way satellite galaxies follow closely the DoS defined by the more luminous dwarfs, thereby further emphasizing the statistical significance of this feature in the Galactic halo. We also notice a deficit of satellite galaxies with Galactocentric distances larger than $100\unit{kpc}$ that are away from the disc-of-satellites of the Milky Way. In the case of Andromeda, we obtain similar results, naturally complementing our previous finding and strengthening the notion that the discs-of-satellites are optical manifestations of a phase-space correlation of satellite galaxies.
\end{abstract}

\begin{keywords}
Galaxies: dwarf, Galaxies: Local Group
\end{keywords}

\section{Introduction}\label{sec_intro}
In recent years, a new class of ultra faint companion galaxies with extremely low stellar densities were detected in the halo of the Milky Way (MW)\footnote{The discovery of the CVn~II dSph was reported twice, by \citeauthor{sakam06} in the issue 653 of ApJ as a letter and in the subsequent issue of ApJ by \citeauthor{belok07}} \citep{willm05a,willm05,sakam06,zucke06c,zucke06,belok06,belok07,walsh07,belok08} by systematically scanning the SDSS-DR6 photometric catalogue \citep{york00}, more than doubling the number of known satellite galaxies within the virial radius of the Milky Way. At the same time, new companions of the Milky Way's sibling, the Andromeda galaxy (M31), were found via deep imaging \citep{marti06,zucke07,majew07,ibata07,irwin08,mccon08}.

The most luminous Milky Way satellite galaxies have been known for more than three decades to exhibit an anisotropic spatial distribution \citep[e.g.][]{lynde76,lynde83,majew94,hartw00}. The quantitative exploration of this phenomenon found that the dwarf galaxies lie close to a virtual plane, the disc-of-satellites (DoS), highly inclined with respect to the stellar disc of the Milky Way \citep*{kroup05,metz07,li08}. Subsequent studies of kinematical data, radial velocities and proper motions \citep{lynde95,palma02}, revealed that a correlation is not only apparent in their three-dimensional distribution, but also as a possible common motion, such that the DoS is likely a rotationally supported structure \citep*{metz08}.

The satellite system of Andromeda shows an asymmetric spatial pattern too \citep{grebe99, hartw00, koch06, majew07}. As for the Milky Way, a disc-of-satellites is evident \citep{metz07}, and the entire satellite system is significantly shifted in the direction of the barycentre of the Local Group \citep{mccon06}. It remains unclear whether this structure is also rotationally supported. Proper motion data are only available for two M31 companions, M33 and IC~10, from VLBI observations of water masers \citep{brunt05, brunt07}, but \citet{vdmar08} concluded that a possible intrinsic rotation of the satellite system is marginal from their statistical analysis. Also no direct proper motion measurement is available for Andromeda itself, but various attempts have been made to constrain the possible transverse motion of M31 based on statistical properties of the motion of its satellite galaxies \citep[e.g.][]{einas82, vdmar08} or the persistence of the stellar disc of M33 under the assumption that it is on a bound orbit to Andromeda \citep{loeb05}.

This paper is to follow up the previous work by \citet{metz07} (MKJ07 hereafter), supplementing the data with more recent findings: the 11+11 newly discovered dSph satellite galaxies of the Milky Way and Andromeda are added to the sample and compared to the proposed disc-of-satellites in Section~\ref{sec_location}. For the Milky Way, a statistic is derived to calculate the probability to find satellite galaxies away from a given reference plane, taking the sky coverage region of SDSS into account, and applied to the most recent discoveries. The results are discussed in Section~\ref{sec_discussion}.

%
\section{Spatial distribution of the new satellites}\label{sec_location}
\subsection{The Milky Way satellites}
\begin{table}
\caption{ Galactocentric coordinates of the seven ultra faint MW satellite galaxies reported in the last three years. 
Four extra objects are listed where the classification as dSph or globular cluster is still uncertain, i.e.~UMa~II, Willman\,1, 
CBe, and Boo~II. In the forth column the orthogonal distances from the DoS are tabulated. See table~1 in MKJ07 for a listing of all the other MW satellites.}
\label{tab_mwdata}
\begin{tabular}{lcccc}
 Name & $l_\oMW$ [$\degr$] & $b_\oMW$ [$\degr$] & $R$ [kpc] & $d_{\rm DoS}$ [kpc] \\
\hline
 UMa~II$^{\rm(a)}$   & 159.6 & +30.0 &  36.5 & 18.5 \\
 Wil~1$^{\rm(b)}$    & 164.7 & +47.7 &  43.0 & 12.7 \\
 CBe$^{\rm(c)}$      & 201.8 & +75.1 &  45.2 &  9.8 \\
 Boo~II$^{\rm(d)}$   & 348.1 & +78.4 &  47.6 & 27.7 \\
 Boo$^{\rm(e)}$      & 356.6 & +77.5 &  57.6 & 32.0 \\
 UMa$^{\rm(f)}$      & 162.0 & +50.8 & 104.9 & 38.3 \\
 Her$^{\rm(c)}$      &  30.9 & +38.8 & 134.2 & 87.2	\\
 CVn~II$^{\rm(c,g)}$ & 132.7 & +80.9 & 150.7 & 19.8 \\
 Leo~IV$^{\rm(c)}$   & 260.0 & +56.2 & 160.6 & 56.7 \\ 
 Leo~V$^{\rm(h)}$    & 256.8 & +58.1 & 180.8 & 57.3 \\
 CVn$^{\rm(i)}$    &  86.9 & +80.2& 219.8 & 43.4 \\
\hline
\end{tabular}
{\footnotesize References: %
$^{\rm(a)}$\,\citet{zucke06c}; 
$^{\rm(b)}$\,\citet{willm05a}; 
$^{\rm(c)}$\,\citet{belok06}; 
$^{\rm(d)}$\,\citet{walsh07}; 
$^{\rm(e)}$\,\citet{belok06}; 
$^{\rm(f)}$\,\citet{willm05}; 
$^{\rm(g)}$\,\citet{sakam06}; 
$^{\rm(h)}$\,\citet{belok08}; 
$^{\rm(i)}$\,\citet{zucke06}
}
\end{table}

\begin{figure}
 \resizebox{\hsize}{!}{
   \includegraphics{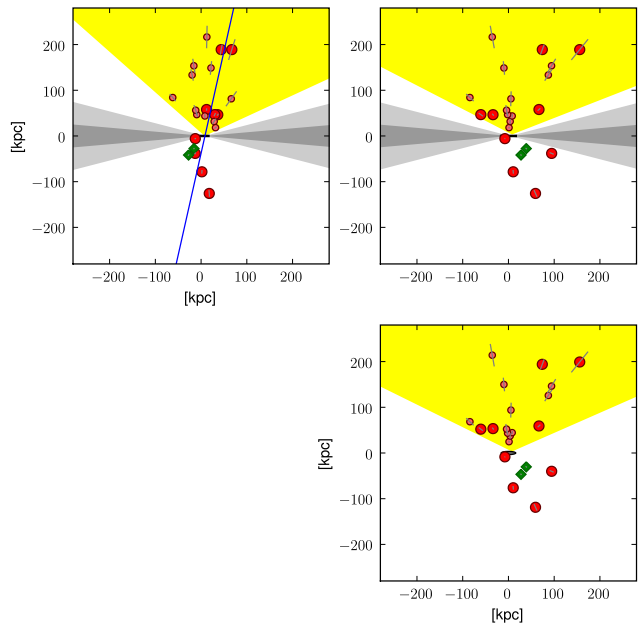}
 }
 \caption{The 3D distribution of the Milky Way satellite galaxies. In the top-left panel an edge-on view onto the fitted disc-of-satellites as given in MKJ07 is shown, derived using the large circles and diamonds. The MW disc, located in the centre of the plot, is seen edge-on. In the top-right panel, a view rotated by $90\degr$ about the polar axis of the Milky Way is shown, and in the lower-right panel a face-on view onto the fitted disc is plotted. The Magellanic Clouds are marked by diamond symbols, the dwarf spheroidals by circles, whereby the smaller circles mark the newly discovered satellites (Table~\ref{tab_mwdata}). Uncertainties are indicated by light grey sticks. In addition, the obscuration region, $|b|<5\degr$, of the MW is shown as the dark-shaded region (the light-shaded region being the $15\degr$ obscuration region). The \emph{projected} northern sky coverage region of the SDSS is indicated by the yellow coloured area.}
 \label{fig_mwsats}
\end{figure}

In MKJ07 we showed that the satellite galaxies of the Milky Way are arranged in a pronounced disc-like distribution, the disc-of-satellites. The direction of the normal vector of the disc is $l_{\oMW}=157.3\degr$, $b_{\oMW}=-12.7\degr$, i.e.~ the DoS is almost perpendicular to the Galactic plane and the closest distance from the Galactic centre is $D_{\rm P}=8.3\unit{kpc}$. The derived rms-height of the disc is $\Delta=18.5\unit{kpc}$, such that $\Delta/R_{\rm vir}=0.15$ is the relative thickness of the disc, where $R_{\rm vir}=250\unit{kpc}$ is the approximate virial radius of the Milky Way. The analysis was carried out for the ``classical'' eleven brightest satellite galaxies. The different projections of the 3-D distribution of the Milky Way satellites are shown in Figure~\ref{fig_mwsats} relative to the DoS. All objects recently discovered in SDSS (Table~\ref{tab_mwdata}) were added to the plot as smaller circles. The projected northern sky coverage region of the SDSS is highlighted by the shaded (yellow) area.

All but one of the newly reported satellite galaxies of the Milky Way are found close to the DoS (Table~\ref{tab_mwdata}). The Hercules (Her) dwarf spheroidal, the left-most data-point marked by a small circle in the top-left panel of Figure~\ref{fig_mwsats}, is $87 \pm 8 \unit{kpc}$ away from the DoS at a Galactocentric distance of $134.2\unit{kpc}$. Without Her, the improved fitting results are basically identical to the MKJ07 fit: $(l_{\oMW}, b_{\oMW}) = (159.7\degr,-6.8\degr) \pm 2.3\degr$, $D_{\rm P}=6.7 \pm 1.3 \unit{kpc}$, and $\Delta=24.9 \pm 1.1 \unit{kpc}$, $1\sigma$ uncertainties were calculated using a Monte-Carlo method. Including Her, the rms-thickness increases to $\Delta=28.5 \pm 1.2 \unit{kpc}$ and the direction of the normal of the fitted plane is $(l_{\oMW},b_{\oMW}) = (149.6\degr, -5.3\degr) \pm 2.2\degr$.

\begin{figure}
 \resizebox{\hsize}{!}{
   \includegraphics{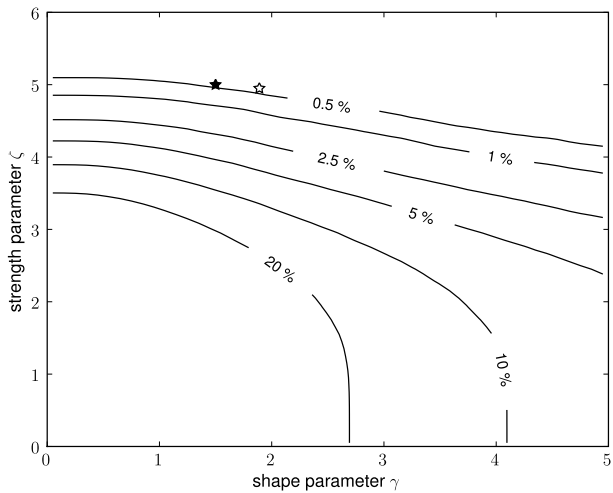}
 }
 \caption{Samples of satellite galaxies were constructed from a spherical isotropic random distribution taking into account i) the obscuration due to the Milky Way disc and ii) the influence of the SDSS sky coverage area. The strength parameter $\gamma$ and shape parameter $\zeta$ were derived for 100\,000 random samples each individually bootstrapped 10\,000 times (see text for a definition of $\gamma$ and $\zeta$). The plot shows contour lines of enclosed values with significance levels $\alpha$ as labelled. The filled star marks the value for all MW satellites, the open star excluding the Hercules dwarf galaxy.
The null-hypothesis that the MW satellites are drawn from a random sample is rejected at the $(1-\alpha)=99.5\%$ confidence level.}
 \label{fig_mwbootstrap}
\end{figure}

The bootstrap analysis described in detail in MKJ07 is repeated, now also incorporating the additional 11 new MW satellites. The resulting shape and strength parameter are $\gamma=1.5$, $\zeta=5.0$. These two parameters describe the distribution of normals fitted to the bootstrapped samples on the sky: the higher the shape parameter $\gamma$, the more symmetric the distribution, and higher values of the strength parameter $\zeta$ indicate more concentrated distributions. 
If Her is excluded from the bootstrapping we find $\gamma=1.9$, $\zeta=5.0$. Even though Her is well away from the DoS, the result of the bootstrapping analysis remains unaffected. The values are shown in Figure~\ref{fig_mwbootstrap} where the confidence contours are plotted for the null-hypothesis that the satellite galaxies were drawn from an isotropic parent distribution. The random samples are setup to resemble the observed satellite galaxies of the Milky Way: what is new in our approach here is that we take into account two observational constraints: (i) the Zone of Avoidance is accounted for by not allowing any satellite to have Galactic latitude $|b| < 15\degr$. The 11 classical satellite galaxies are setup with this bias only. (ii) The SDSS sky coverage is accounted for by allowing only satellites with $b > 30\degr$ for the remaining ones. The null-hypothesis is rejected at the $(1-\alpha)=99.5\%$ confidence level. This clearly shows that, even though the updated DoS is somewhat ``thicker'' than for the classical satellites alone, the currently known Milky Way satellite galaxies are highly unlikely drawn from an isotropic distribution (cf.\ MKJ07).

\subsection{The effect of the SDSS sky coverage}
The SDSS survey area covers about 20 percent of the total sky, being merely a spherical cap at the Galactic North pole with $b\gtrsim +30\degr$, except for three narrow stripes in the southern Galactic sky. Certainly, this observational bias must have an influence on the derived spatial distribution of the satellite galaxies discovered in the SDSS data. Especially, since the DoS is almost perpendicular to the MW plane, the probability of finding satellite galaxies in the polar region close to the DoS is higher than at low Galactic latitudes.

One can now ask what the probability is to find a satellite galaxy at an orthogonal distance $d$ from the DoS. For any given fixed plane at a distance $0 \le D_P < R$ from the Galactic centre, the probability to find a satellite galaxy at a distance $d$ perpendicular to this plane for a spherical isotropic parent distribution can be analytically expressed. The derived distribution function (DF), $F_R(d)$, is \emph{independent} of the radial density distribution function of the satellite galaxies.
For a cone-like region ($b>b'$) about an axis parallel to the reference-plane, as in the case of SDSS, the distribution function $F_{R,b>b'}(d)$ can also be written down, but needs numerical integration. An important point to note is that the DF rises more steeply for the cone coverage area than for a full-sky (or full hemisphere) coverage area. This means that it is more likely to discover a satellite galaxy close to the DoS when only a polar cap is observed as is done for the SDSS. For example, at $R=60\unit{kpc}$ the probability to find an object at a distance $d=25\unit{kpc}$ from the DoS increases by 13 percent from $F_{R=60}(25)=0.42$ to $F_{R=60,b>+30}(25)=0.55$.

The derived distribution function for a cone region is strictly speaking correct only for a plane parallel to the axis of symmetry. The fact that the DoS is tilted with respect to this axis by $12.7 \degr$ introduces a correction factor of the order of 2 percent, $\cos(12.7\degr) \approx 0.98$, that will be neglected here. Also, the SDSS coverage region is not a perfect cone symmetric with respect to the Galactic North Pole (compare also figure~1 in MKJ07), but only approximately so.

\begin{figure}
 \resizebox{\hsize}{!}{
   \includegraphics{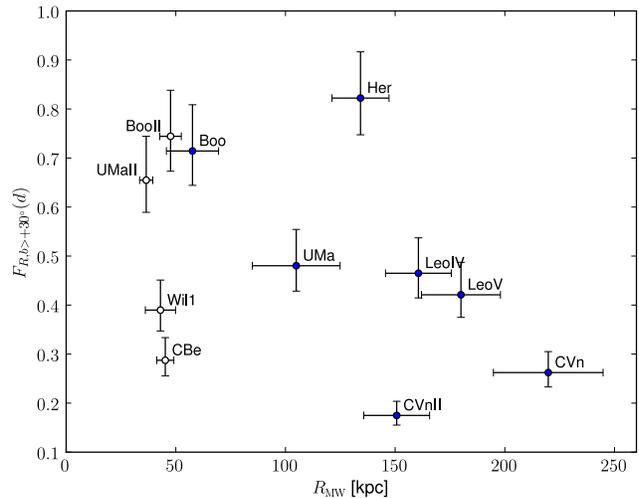}
 }
 \caption{The probability that a satellite galaxy is found in the SDSS at its derived distance $d$ or closer to the DoS versus their Galactocentric distance $R$. Galactocentric distance uncertainties were calculated from Heliocentric distance uncertainties as taken from the literature. Uncertainties in the $F_R(d)$ parameter were computed by considering sky coverage areas of $b>20\degr$ and $b>40\degr$, respectively. Open symbols mark objects where the classification as a dSph or disrupted globular cluster is uncertain.}
 \label{fig_r_pd}
\end{figure}

\bigskip
The probability that a satellite is found in the SDSS catalogue at its derived distance $d$ or closer to the DoS is plotted in Figure~\ref{fig_r_pd} for $D_P=3.3\unit{kpc}$, which is the offset of the DoS from the Sun (as the Sun is at the vertex of the cone-like SDSS sky coverage region).
For an isotropic distribution these values are expected to be uniformly distributed. For objects with $R<100\unit{kpc}$ the probabilities are consistent with a uniform distribution, but for the new dwarf spheroidals with $R>100\unit{kpc}$ only Hercules is found in the upper region. In contrast, five of the outer dSph have $F<0.5$.

It is important to recall that $F_{R,b}(d)$ describes a property of the sky coverage area and does not tell anything about the true spatial distribution of the satellite galaxies. Consider, for example, a sky coverage region of only $10\degr$ about the north pole. All satellite galaxies found in such a region must be very close to a polar DoS independent of whether they are highly an-isotropically distributed or not, and the DF may easily have values of order one. Nevertheless, from Figure~\ref{fig_r_pd} we can conclude that there is a deficit of satellite galaxies far off the DoS at galactocentric distances $R>100\unit{kpc}$.

It is worthwhile to consider that those objects marked with open symbols are likely strongly influenced by tides \citep{gilmo08,vdber08}. If they came close to the Galactic disc, they are also likely prone to precession in the non-spherical potential induced by the Galactic disc or to scattering events \citep{zhao98,penar02}. Only for the B\"ootes dwarf at $R=57.6\unit{kpc}$ there seems to be a general agreement that it has to be classified as a dSph; its total luminosity is about ten times higher than for all the other objects with $R<100\unit{kpc}$. If the other four objects, UMa~II, Willman\,1, CBe, and Boo~II, are not dSphs they may also have a completely different parent distribution. Interestingly, also the Her dSph has been strongly affected by tides \citep{colem07}. If this is due to a recent close perigalactic passage, the orbit of Her might have been influenced by the Milky Way disc, thus possibly causing a drift of the orbit out of the DoS if it were within before.

\subsection{The Andromeda galaxy}
\begin{figure}
 \resizebox{\hsize}{!}{
   \includegraphics{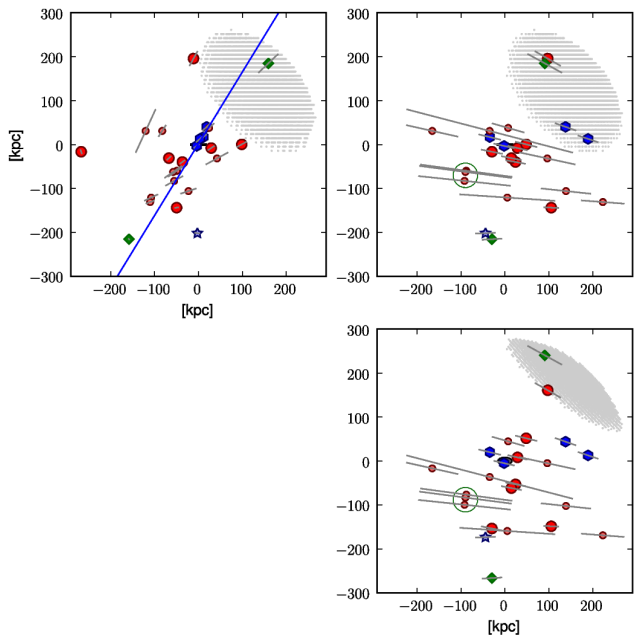}
 }
 \caption{The 3D distribution of Andromeda satellite galaxies as in Figure~\ref{fig_mwsats} for the Milky Way. Galaxy types are indicated as follows: dIrr galaxies are marked by diamonds, dEs by hexagons, and dSphs by circles, smaller circles marking the newly discovered dSphs. The location of M33 is marked by the star symbol. The light grey sticks are distance one-sigma uncertainties of the satellites only. The grey shaded area indicates the projected region where potentially some satellite detections may be hindered by foreground MW structure \citep[see also][]{mccon06}. A group of satellite galaxies, consisting of And~XI, XII, and XIII, is marked by a light circle. These dSph were assumed to all have the same heliocentric distance.}
 \label{fig_m31sats}
\end{figure}
For the Andromeda satellites we performed the same steps as for the Milky Way: The recently discovered companions listed in Table~\ref{tab_m31data} were added to the sample and are plotted in Figure~\ref{fig_m31sats} with respect to the Andromeda DoS as given in MKJ07. For this figure the data-set taken from \citet{mccon06} is used and supplemented with data as given in Table~\ref{tab_m31data}. Again, most of the recently discovered satellite galaxies are found close to the previously fitted disc-of-satellites. This is particularly remarkable as the Andromeda system is much less affected by foreground obscuration due to the disc of the Milky Way \citep{mccon06}, and it is also not as biased as is the Milky Way system towards the Galactic north pole region. Only the two Andromeda companions reported very recently \citep{mccon08} lie off the disc, whereas the third one, And~XVIII at $R_\oAnd=589\unit{kpc}$ and thus outside the virial radius of Andromeda, is again found close to the M31 DoS. As for the Milky Way, incorporating the data of all the newly discovered dSphs to the fitting methodology of MKJ07 has only a marginal influence on the fitting parameters: $l_{\oAnd}=60.2\degr$, $b_{\oAnd}=-30.7\degr$, $D_{\rm P}=15.6\unit{kpc}$, and $\Delta=45 \unit{kpc}$, whereby the strongest effect is caused by those two outlying satellites, And~XIX and XX. Without them we find $l_{\oAnd}=70.2\degr$, $b_{\oAnd}=-32.9\degr$, $D_{\rm P}=1.7\unit{kpc}$, and $\Delta=39.2 \unit{kpc}$, compared to the results as previously derived in MKJ07: $l_{\oAnd}=73.4\degr$, $b_{\oAnd}=-31.5\degr$, $D_{\rm P}=1.0\unit{kpc}$, and $\Delta=45.9 \unit{kpc}$ ($l_{\oAnd}$ and $b_{\oAnd}$ are the longitudes and latitudes of the pole of the fitted DoS in the Andormeda centric coordinate system as defined in MKJ07).

The proximity of the dwarf spheroidals to the M31 DoS is in particular not strongly influenced by the still uncertain distances to them which are of the order 10\%, as the DoS of M31 is seen nearly edge-on from the Sun. Thus, a large distance uncertainty only marginally influences their perpendicular distance to the DoS (see the indicated line-of-sight distance uncertainties in the top-left panel of Figure~\ref{fig_m31sats}). Thus also the apparent grouping of And~XI, XII, and XIII, marked by the light circle, which is caused by their assumed common heliocentric distance \citep{marti06}, is not biasing the finding that these satellites are close to the M31 DoS.

\begin{table}
\caption{Positions of the recently discovered Andromeda dSph satellite galaxies in Andromeda-cetric coordinates as defined in MKJ07 and their distances to the M31 DoS. Heliocentric coordinates and distances were taken from the respective discovery works, the assumed distance to Andromeda used for the transformation is $785\unit{kpc}$ \citep{mccon06}. And~XI -- XIII were all assumed to be located at a common Heliocentric distance and thus appear to be clustered (see also Figure~\ref{fig_m31sats}), which may in reality not be valid. We also list And~XVIII here for completeness, but with a distance of $589\unit{kpc}$ this galaxy is well outside the virial radius of M31.} \label{tab_m31data}
\begin{tabular}{lddcc}
 Name &
   \multicolumn{1}{c}{$l_\oAnd$\,[$\degr$]} &
   \multicolumn{1}{c}{$b_\oAnd$\,[$\degr$]} &
   \multicolumn{1}{c}{$R_\oAnd$\,[kpc]}     &
   \multicolumn{1}{c}{$d_{\rm DoS}$\,[kpc]}\\
\hline
 And~XVII$^{\rm(a)}$  &  90.6 &  56.5 &  45 &  4\\
 And~X$^{\rm(b)}$     & 139.8 & -16.7 & 110 & 46\\
 And~XII$^{\rm(c)}$   & 314.2 & -30.6 & 117 &  3\\
 And~XI$^{\rm(c)}$    & 310.8 & -30.3 & 124 &  8\\
 And~XX$^{\rm(f)}$    & 269.5 &  13.8 & 129 & 116\\
 And~XIII$^{\rm(c)}$  & 312.3 & -37.6 & 136 &  5\\
 And~XIV$^{\rm(d)}$   & 250.6 & -48.2 & 162 & 20\\
 And~XV$^{\rm(e)}$    & 172.8 & -36.9 & 176 & 33\\
 And~XIX$^{\rm(f)}$   & 316.9 &   9.5 & 188 & 79\\
 And~XVI$^{\rm(e)}$   & 189.7 & -27.8 & 281 & 27\\
 And~XVIII$^{\rm(f)}$ & 356.9 &  31.7 & 589 & 61\\
\hline
\end{tabular}
{\footnotesize References: %
$^{\rm(a)}$\,\citet{irwin08}; 
$^{\rm(b)}$\,\citet{zucke07}; 
$^{\rm(c)}$\,\citet{marti06}; 
$^{\rm(d)}$\,\citet{majew07}; 
$^{\rm(e)}$\,\citet{ibata07}; 
$^{\rm(f)}$\,\citet{mccon08}
}
\end{table}

\section{Concluding remarks}\label{sec_discussion}
In Figure~\ref{fig_mwsats}, the biasing due to the sky-coverage area of the SDSS is obvious. Nevertheless, there are remarkably large portions of the sky that are way off the DoS -- but no satellite galaxies are reported to be located there (top-left panel in Fig.~\ref{fig_mwsats}). Especially at large Galactocentric distances, $R>100\unit{kpc}$, there is a deficit of satellites far off the DoS (Fig.~\ref{fig_r_pd}). The same is true for the Andromeda system, which is much less influenced by obscuration: only the two most recently reported Andromeda companions \citep{mccon08} are not found close to the disc, but eight are found very close to the DoS. \citet{majew07} noted that along a radial vector from the centre of M31, connecting NGC~147 and And~XIV, in total six satellite galaxies are located in projection, and \citet{irwin08} found that And~XVII is also on that line. We demonstrated that the new dSphs lie on a straight line not only in projection, but taking their full three-dimensional data into account, we find that they belong to the same disc-of-satellites as reported in our previous work (MKJ07). This holds true despite large distance uncertainties.

It appears highly unlikely that the disc-of-satellites for the two dominant galaxies in the Local Group are the result of observational biases.
The bootstrapping analysis shows, taking the biasing effects due to the SDSS sky coverage area into account, that it can be excluded at a very high significance level of 99.5\% that the MW satellites are drawn from an isotropic distribution, even then when we include the outlying satellite Hercules. At the same confidence level it has been excluded that the classical 11 MW satellites are drawn from an isotropic distribution \citep{metz07}. The findings presented here strengthen the case that the disc-of-satellites for both, the Milky Way and Andromeda, are real and evidence of a spatial \citep[and kinematical:][]{lynde95,palma02,metz08} correlation of the satellite galaxies.
We emphasise that it has been found that the orbital angular momenta of the innermost classical satellites point in a direction close to the pole of the DoS \citep{metz08}, which itself is mostly defined by the outer satellites, now including also the new discoveries. This suggests a strong phase-space correlation of the satellites.

Different solutions have been proposed to account for the satellite galaxies of the Milky Way in the context of them being cold-dark matter (CDM) sub-structures. A strong central clustering of sub-haloes that mimics a ``thin'' DoS distribution \citep{kang05,zentn05} is unlikely to be the explanation (MKJ07). The most massive sub-haloes before accretion \citep{libes05} do not account for the apparent rotational support of the Milky Way DoS \citep{metz08}. Other solutions, not directly addressing the DoS problem, range from the most massive sub-haloes \citep{stoeh02} to the least massive sub-haloes \citep{sales07a}, or the earliest forming sub-haloes \citep{strig07} as being the Local Group satellites. Recently, it has been proposed that infalling groups of sub-haloes can account for the satellite galaxies \citep{li08,dongh08} or that the dwarfs form in chain structures tracing the dark matter filaments \citep{ricot08}. This would imply that most satellite galaxies belong to a single group that came in, or that multiple groups came in from the same direction along the intermediate-scale filamentary structure. But this direction must then have been different for the Milky Way and Andromeda, and can not have been the supergalactic plane. Not all of the proposed scenarios to explain the satellite galaxies as luminous dark matter dominated haloes mutually exclude each other, but it appears to be very controversial within the CDM community which are the main characteristics that make dark matter sub-haloes to be luminous.

A possible alternative is that most of the satellite galaxies are of tidal origin \citep{zwick56,lynde83}. If the interaction of a gas-rich galaxy with the proto Milky Way took place in the early Universe, a large number of tidal dwarfs would have likely been produced \citep{okaza00}, that are naturally arranged in a disc-like structure: due to the conservation of angular momenta, their orbital poles are determined by the plane of the interaction. This process may also explain dwarf galaxies on similar orbits if they were formed in adjacent regions in the tails \citep{belok08}. Initially gas-rich, the TDGs may retain their gas for some time, having prolonged star formation epochs \citep{recch07}, and may evolve into dSph like galaxies we see today \citep{kroup97,metz07b}. But it remains to be seen whether Newtonian dynamics can account for the detailed velocity distribution functions observed in the satellites \citep{lokas01,genti07}.

Recently, looking at distant host galaxies in the SDSS, \citet{baili08} and \citet{steff08} found for isolated blue galaxies that their satellites are, in projection, isotropically distributed  (in contrast to what \citealp{holmb69} originally found). The MW and M31 seem to be exceptions in this case as both have highly an-isotropic satellite distributions. It is, however, worthwhile to consider that per host typically only one single satellite is found in these studies. All of the dSphs, which build up the DoS, would not be considered in these analyses due to their faintness. That said, from a statistical point of view the brightest satellites (LMC or M33 type) can be isotropically distributed for a large sample of hosts. But each primary individually might well have a highly an-isotropically distributed satellite galaxy system of dSphs.

The next years will be exciting: new satellite galaxies are expected to be discovered by upcoming search campaigns such as the Stromlo Missing Satellites Survey \citep{jerje08} -- especially also at lower Galactic latitudes. If the expected ultra faint galaxies turn out to be isotropically distributed, we are still left with the fact that the most luminous MW satellites are arranged in the disc-of-satellites, whereas the low luminous systems would be isotropically distributed. However, if they turn out to be located within the disc-of-satellites, this will substantiate the suggested strong correlation of the satellite galaxies, and it will become more challenging to explain them as luminous, dark matter dominated sub-structures. It is quite clear that dSph satellite galaxies hold important clues not only to the physical processes acting at early cosmological times, but also on the foundation of cosmological theory.

\vspace{5mm}
\noindent{\bf Acknowledgements}\\
{The authors acknowledge financial support from the Go8/DAAD Australia Germany Joint Research Co-operative Scheme.}

\bibliographystyle{aa_mn2e}
\bibliography{quotes,quotebooks,quotes3,quotes2,quotes_preprint}

\label{lastpage}
\end{document}